\documentclass[twocolumn]{aastex631}

\usepackage{amsmath}
\usepackage{xspace,xcolor}
\usepackage{CJK}

\newcommand{\chandra}{{\it Chandra}\xspace}
\newcommand{\msun}{M$_{\sun}$\xspace}
\newcommand{\nh}{$N_{\rm H}$\xspace}
\newcommand{\teff}{T_{\rm eff}}

\shorttitle{Cas A cooling with Chandra HRC-S}
\shortauthors{Zhao et al.}

\begin{document}
\begin{CJK*}{UTF8}{gbsn}

\title{Verification of Cas A neutron star cooling rate using Chandra HRC-S observations}

\author[0000-0002-7716-1166]{Jiaqi Zhao (赵嘉琦)}
\affiliation{Physics Department, CCIS 4-183, University of Alberta, Edmonton, AB, T6G 2E1, Canada}

\author[0000-0003-3944-6109]{Craig O. Heinke}
\affiliation{Physics Department, CCIS 4-183, University of Alberta, Edmonton, AB, T6G 2E1, Canada}

\author[0000-0002-5810-668X]{Peter S. Shternin}
\affiliation{Independent researcher}

\author[0000-0002-6089-6836]{Wynn C. G. Ho}
\affiliation{Department of Physics and Astronomy, Haverford College, 370 Lancaster Avenue, Haverford, PA 19041, USA}

\author[0000-0003-2596-196X]{Dmitry D. Ofengeim}
\affiliation{Racah Institute of Physics, The Hebrew University, Jerusalem 91904, Israel}

\author[0000-0002-7507-8115]{Daniel Patnaude}
\affiliation{Smithsonian Astrophysical Observatory, 60 Garden Street, Cambridge, MA 02138, USA}

\correspondingauthor{Jiaqi Zhao, Craig O. Heinke}
\email{jzhao11@ualberta.ca, heinke@ualberta.ca}



\begin{abstract}

The young neutron star (NS) in the Cassiopeia A (Cas A) supernova remnant is a fascinating test for theories of NS cooling. Chandra observations have indicated that its surface temperature is declining rapidly, about 2\% per decade, using 20 years of data, if a uniform carbon atmosphere is assumed for the NS. This rapid decline may be caused by the neutrons in the NS core transitioning from a normal to a superfluid state. However, most of the Cas A NS observations were performed by the Chandra ACIS detectors, which suffer complicated systematic effects. Here, we test the cooling of the Cas A NS with Chandra HRC data over 25 years. The Chandra HRC detector has independent systematics, serving as a cross-check. Assuming a fixed hydrogen column density (\nh), we infer 
the cooling rate of the Cas A NS to be {0.57$^{+0.26}_{-0.27}$\%} per decade. 
Allowing the \nh to vary with time (as estimated using ACIS data), the cooling rate is {1.11$^{+0.25}_{-0.28}$\%} per decade. These {cooling} rates are smaller than measured using ACIS data, implying systematic uncertainties have not been eradicated from either or both datasets. However, 
we have verified the decline in the absorbed flux from the Cas A NS using an independent instrument, at $>3\sigma$ level {($4.7\% \pm 1.5\%$ over 10 years)}. 
Additionally, the weaker cooling rate of Cas A NS inferred from HRC datasets eliminates the tension with the theoretically predicted cooling, and can be explained by the reduced efficiency of the
neutrino emission accompanying the Cooper pair breaking and formation process in neutron triplet-state superfluid.

\end{abstract}



\section{Introduction} \label{sec:intro}

Cassiopeia A (Cas A) is a supernova remnant (SNR) located in the constellation of Cassiopeia at a distance of {3.3~kpc \citep{Reed1995,Alarie2014}}, with an age of about 340 years \citep{Fesen2006}.
The central compact object (CCO) in Cas A, the youngest known neutron star (NS; hereafter Cas A NS), was discovered through the \chandra first-light observation in 1999 \citep{Tananbaum1999}.
No pulsations have been detected from this CCO in either X-rays \citep{Murray2002,Ransom2002,Halpern2010} or radio \citep{McLaughlin2001}, {with a reported upper limit on pulsed fraction of 11\% at 99\% confidence \citep{Alford2023}}.
The absence of X-ray pulsations 
{was interpreted as evidence that} the X-ray emission originates from the entire object's surface.
X-ray spectral analysis is consistent with a NS with a uniform carbon atmosphere and a low surface magnetic field strength \citep{Ho2009,Chang2010,Wijngaarden2019,Ho2021}.
However, alternative models 
{such as an isotropic blackbody model with small hotspots and small viewing and hotspot inclination angles \citep{Wu2021,Alford2023}}
might also explain the X-ray emission from Cas A NS.
{Intriguingly, assuming a carbon atmosphere model,} \citet{Heinke2010} analyzed archival \chandra Advanced CCD Imaging Spectrometer (ACIS) data (Graded mode) from 2000 to 2009, and reported a decline in the surface temperature of the Cas A NS by 4\%, from $(2.12\pm0.01)\times10^6$~K to $(2.04\pm0.01)\times10^6$~K (however, see below). 
The Cas A NS marked the 
first evidence for 
a young, {isolated} non-magnetar neutron star having a real-time measurement of a {non-zero} temperature derivative, and as of now, the Cas A NS remains the only one with such measurements, despite numerous neutron stars having recorded measurements of temperatures 
\citep[e.g.][]{Ho2021}. 

The cooling of neutron stars principally results from neutrino emission from their cores, and most of the cooling processes operate throughout the neutron star lifetime \citep{Yakovlev2004}.
However, the drastic cooling rate found in the Cas A NS is not likely to be produced by typical cooling processes, like the modified URCA process \citep{Heinke2010,Yakovlev2011}.
The observed cooling of the Cas A NS can be successfully interpreted if the neutrons in the NS core are in a superfluid state, producing enhanced neutrino emission from Cooper pair breaking and formation (PBF; \citealt{Flowers1976}), leading to a rapid decline of the surface temperature \citep{Page2011,Shternin2011}.
Thus, measurements of the cooling rate may 
constrain the properties of the NS core.
For example, \citet{Shternin2021} constrained the maximal critical temperature of neutron pairing in the Cas A NS core to the range of $[5,10]\times 10^8$~K, using an {equation of state and superfluid} model-independent analysis (see also \citealt{Shternin2023}).
Alternative explanations of the rapid cooling have also been proposed, such as $r$-mode damping process \citep{Yang2011}, dissipation of magnetic field \citep{Bonanno2014}, 
{delayed crust-core relaxation due to in-medium suppression of the thermal conductivity of the core \citep{Blaschke2012}, direct Urca cooling in a tiny central region \citep{Leinson2022}, {joint effect of Urca (direct Urca plus modified Urca) processes \citep{Potekhin2025}}, etc.} 

An issue is that the initial ACIS observations of the Cas A NS temperature decline suffer several systematic effects, that are hard to calibrate.
The Cas A NS is bright enough that multiple photons can land on the detector during a single readout time when reading the full detector, inducing ``pileup'', where multiple photons are erroneously read as a single (higher-energy) photon, or mistakenly identified as a cosmic ray and discarded.
In addition, the ACIS detector has suffered an increasing buildup of an organic molecular contaminant over time, which changes the effective area of the telescope, and makes calibration complicated\footnote{\url{https://cxc.cfa.harvard.edu/ciao/why/acisqecontamN0015.html}} \citep{Plucinsky2022}. 
Most observations of Cas A are taken in an instrument mode that reads out the entire chip (thus imaging the full SNR), but the high count rate would exceed the satellite telemetry limits, so these observations only telemeter reduced information about each detected photon (``GRADED'' mode).
Using GRADED mode makes it harder to correct for pileup, especially with the changing detector effective area. 

Several subsequent \chandra observations and studies that attempted to avoid {some of} these systematic effects {(eliminating pileup and telemetry saturation by reading out only a subarray)} showed significantly less, or no, cooling of the Cas A NS, compared to the cooling rate reported by \citet{Heinke2010}, introducing doubt about the rapid cooling of the Cas A NS. 
\citet{Elshamouty2013} analyzed \chandra High Resolution Camera (HRC) data, which has at least different systematics compared to the ACIS detector, and obtained a cooling rate of only 1.0\%$\pm$0.7\% per decade. 
Moreover, \citet{Posselt2013} used two ACIS datasets taken using a subarray (reading out only part of the CCD), which 
mostly removes 
 pileup effects, in Faint mode (telemetering more information on each detected photon), and found no statistically significant temperature decline.
{\citet{Posselt2018} further placed a 3$\sigma$ upper limit on the cooling rate of $<$3.3\% per decade with an additional set of observations in the same instrument mode.}
On the other hand, 
{multiple full-frame ACIS GRADED observations of the Cas A NS over 20 years, with progressive}
improvement of instrument calibration \citep[see, e.g.,][]{Plucinsky2022}, have continued to {support} cooling, but with a reduced {10-year} rate of 2-3\% 
\citep{Wijngaarden2019,Ho2021}.
Recently, the two independent cooling rate measurements for the Cas A NS 
have converged, 
with ACIS full-frame Graded measurements giving 1.6-2.2\% per decade \citep{Shternin2023}, and ACIS subarray FAINT measurements giving 1.5-2.3\% per decade \citep{Posselt2022}.
Thus, it seems the evidence for the rapid cooling of the Cas A NS is becoming {stronger}, though the cooling rate appears to be about {half of the rate first measured with only 9 years of data.}

However, since the majority of \chandra observations of Cas A were performed with ACIS detectors (in either GRADED or subarray FAINT mode), the measurements of temperature decline by the two groups are not completely independent. Uncertainties in how the ACIS effective area has changed over time can still affect both measurements. 
HRC observations, however, can provide an independent verification of the cooling rate of the Cas A NS, 
as the HRC detectors are not affected by contamination and pileup, 
though they do suffer 
from a steady decline with weak wavelength dependence
in quantum efficiency.\footnote{\url{https://cxc.harvard.edu/proposer/POG/html/chap7.html}}
The HRC-S datasets used by \citet{Elshamouty2013} were taken from 1999 to 2009, 
giving a 
measurement of the cooling rate (1.0\%$\pm$0.7\% per decade) that was {only somewhat} significant.
In this paper, we 
measure the cooling of the Cas A NS using newly obtained HRC-S observations in November 2023 and August 2024 (PI: Craig Heinke) as well as previously archived HRC-S datasets. With data taken over a time span of $\sim$25 years (more than twice the time span in \citealt{Elshamouty2013}), we aim to {measure} 
the cooling rate of the Cas A NS with higher significance, and provide an independent check on recent ACIS measurements. 

\section{Observations and data reduction}
\label{sec:observation}

The Cas A remnant was observed by \chandra HRC-S during four previous epochs (1999, 2000, 2001, and 2009), with five new observations (PI: Craig Heinke) conducted recently in November 2023 and August 2024.
We retrieved all the useful HRC-S datasets with an exposure time $\gtrsim$10 ks, leading to a total exposure time of 652.2~ks (Table~\ref{tab:obs}).
Note that three observations (ObsIDs 173, 174, and 1408) were not included in this analysis as the Cas A NS is highly off-axis in those observations (off-axis angles $\theta \gtrsim$ 20\arcmin).
We reprocessed the datasets using {\tt chandra\_repro} script in {\sc ciao}\footnote{Chandra Interactive Analysis of Observations, available at \url{https://cxc.cfa.harvard.edu/ciao/}.} (version 4.16 with {\sc caldb} version 4.11.5).

Due to the poor spectral resolution of HRC-S ($\Delta E/E \sim 1$ at 1 keV), spectral analysis of the Cas A NS is 
not useful. 
{Instead, to find the best-fit effective temperatures ($T_{\rm eff}$) of the Cas A NS, we looked into the total numbers of counts of simulations and observations.}
{Specifically, to model the counts from Cas A NS with different $\teff$}, we simulated the spectra for each observation using the {\tt fake\_pha} script in {\sc sherpa}\footnote{{\sc ciao}'s modeling and fitting package.} (version 4.16.0).
The spectral model used for simulations is the same as in \citet{Elshamouty2013} (see also, e.g., \citealt{Wijngaarden2019,Shternin2023}),
including a non-magnetized carbon atmosphere NS model ({\tt xsnsx} in {\sc sherpa}; \citealt{Ho2009}), the Tuebingen-Boulder interstellar medium absorption model ({\tt xstbabs}) with {\it wilm} abundance \citep{Wilms2000} and {\it vern} cross sections \citep{Verner1996}, and a dust scattering model ({\tt xsspexpcut}; \citealt{Predehl2003}).
{We adopted the best-fit values of the model parameters, except for the NS radius ($R_{\rm NS}$), from \citet{Shternin2023}}:
for the {\tt xsnsx} model, the NS mass ($M_{\rm NS}$) 
was set to be 1.60~\msun 
, while the distance to the Cas A NS was fixed at 3.33~kpc \citep{Reed1995,Alarie2014}; 
the hydrogen column number density (\nh) towards the Cas A NS of the {\tt xstbabs} model was assumed to be fixed at $1.656 \times 10^{22}~{\rm cm}^{-2}$ {(see below for varying \nh)}; for the {\tt xsspexpcut} model, we set the exponent index $\alpha=-2$ and the characteristic energy $E_{\rm cut}=[0.49 N_{\rm H} (10^{22}~{\rm cm}^{-2})]^{1/2}\,{\rm keV}= 0.9$~keV. 
{We set the NS radius to $R_{\rm NS}=12.6$~km which is well within the posterior credible range $R_{\rm NS}=13.7^{+1.1}_{-1.8}$~km from \citet{Shternin2023}, but at the same time is more consistent with the current observational and theoretical constraints on the NS radius 
\citep[e.g.][]{Koehn2025}. As such, this selection allows us to employ a realistic equation of state in the illustrative cooling simulations below.}

\begin{figure*}
    \centering
    \includegraphics[width=\linewidth]{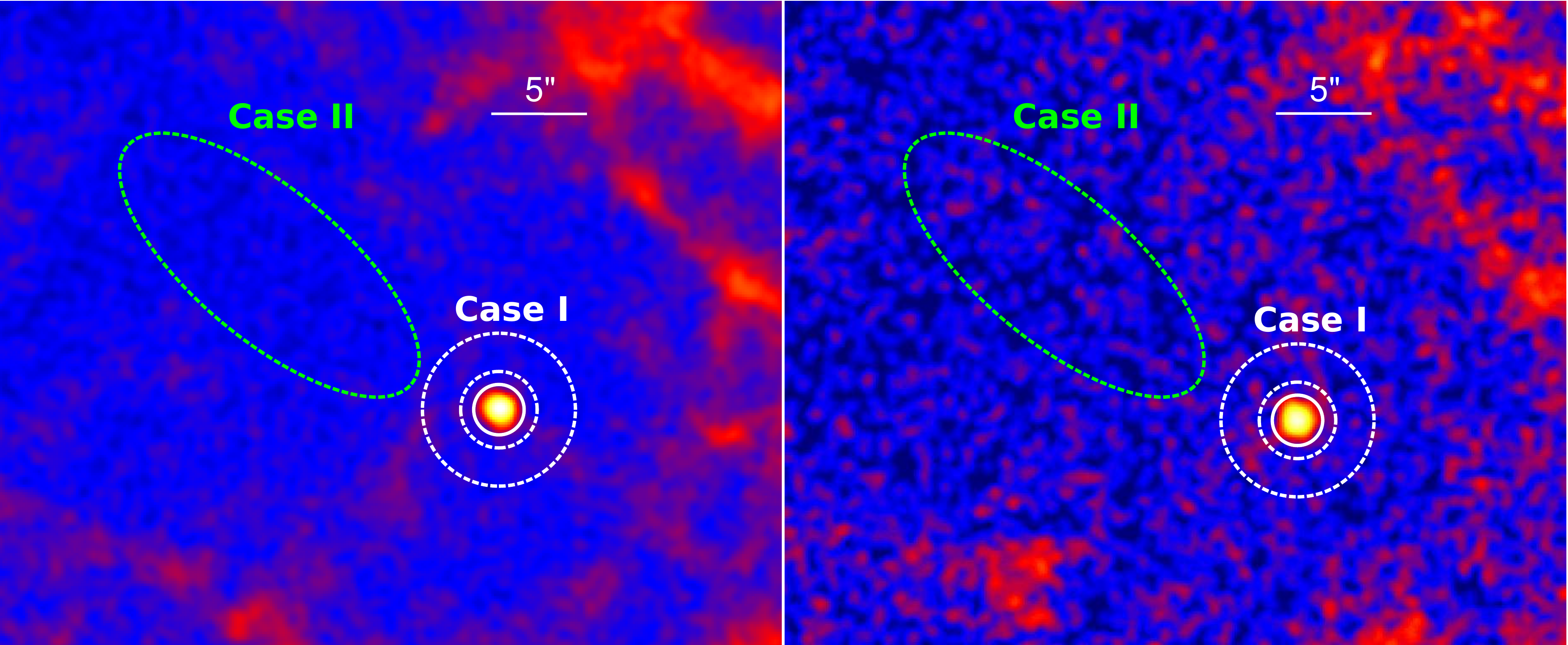}
    \caption{{The exposure-corrected, smoothed images}
    of the Cas A NS taken with Chandra HRC-S in {March 2009 (left; ObsIDs 10227)}
    and in {November 2023 (right; ObsIDs 26629)}.
    In each image, the solid {white} circle shows our extraction region of the Cas A NS ($r_{\rm src}=1.3\arcsec$, or 10 HRC pixels), while the dashed white annulus and green ellipse show the background extraction regions for Case I and Case II, respectively. Significant filaments around the Cas A NS can be seen in the 2009 epoch.}
    \label{fig:combined_image}
\end{figure*}

We used the {\tt specextract} script to extract and generate the instrumental response files, i.e. the effective area files (ARFs) and redistribution matrix files (RMFs), from the corresponding observations.
The source extraction region was defined as a circle with a radius $r_{\rm src}=1.3\arcsec$ (equivalent to 10 HRC pixels) centered on the Cas A NS.
For the background region, we considered two cases. 
Case I background was extracted from an annular region ($r_{\rm bkg} = 2\arcsec-4\arcsec$) centered on the Cas A NS.
However, as mentioned by \citet{Elshamouty2013}, transient filaments around the Cas A NS were found in observations during 2009 (see Figure~\ref{fig:combined_image}), whereas no significant filaments were present around the Cas A NS in other observations.
{To consider alternative background models (depending on whether or not the filament actually is present inside our CCO spectral extraction region),}
we therefore considered a second background extraction region (Case II), which is an ellipse in a low-background area adjacent to Cas A NS (Figure~\ref{fig:combined_image}).
The 
observed count rates were also obtained from the {\tt specextract} output files.
{In addition, we calculated the background-subtracted energy flux for Cas A NS, using {\sc ciao}'s tool {\tt srcflux}, which implements the {\tt eff2evt} tool to compute the 
energy flux with the quantum efficiency and effective area taken into account. 
Note that, since energy is not accurately measured in HRC observations, we selected an effective energy of 1.5~keV to calculate the quantum efficiency and effective area, {which corresponds to the peak of the NS absorbed spectrum}. 
The measured net energy fluxes are listed in Table~\ref{tab:obs}.}

To {infer the}
effective temperature of the Cas A NS for each observation,
{we derived posterior probability distributions with the nested sampling Monte Carlo algorithm
MLFriends \citep{Buchner2016,Buchner2019} using the
UltraNest\footnote{\url{https://johannesbuchner.github.io/UltraNest/}} package \citep{Buchner2021}.
We defined a Poisson likelihood as
\begin{equation}
\begin{split}
    \mathcal{L}(S,B|T_{\rm eff},N_b) & = \frac{(N(T_{\rm eff})+N_b)^S e^{-(N(T_{\rm eff})+N_b)}}{S!} \\
    & \quad \times \frac{(N_b f)^Be^{-N_b f}}{B!},\\
    \ln{\mathcal{L}} & = S \ln{(N(T_{\rm eff}) + N_b)} + B \ln{(N_b f)} \\
    & \quad - (N(T_{\rm eff}) + N_b + N_b f) + {\rm const},
\end{split}
\end{equation}
where $S$ and $B$ are the observed total counts from the source and background regions, respectively; $N(T_{\rm eff})$ is the modeled count number from Cas A NS as a function of $\teff$, while $N_b$ is the modeled background count from the source region; $f$ is the ratio of the background to source region sizes {($f=$ 6.9 and 21.7 for Case I and II, respectively)}; model-independent terms {in log-likelihood} are absorbed into a constant that is omitted in the fitting procedure.
We employed uniform priors for $\log\teff$ and $N_b$ in the range [6.0,6.5] and [0, 1000], respectively.}
The 
{inferred} $\teff$ values are shown in Table~\ref{tab:Teff}.

\begin{splitdeluxetable*}{lccDCBlCCC|CCC}

\tabletypesize{\footnotesize}

\tablecaption{\chandra HRC-S observations used in this work and the corresponding inferred count rates and fluxes of the Cas~A NS. \label{tab:obs}}

\tablenum{1}


\tablehead{\colhead{} & \colhead{} & \colhead{} & \multicolumn2c{} & \colhead{} & \colhead{} & \multicolumn3c{Case I} & \multicolumn3c{Case II} \\
\cline{8-13}
\colhead{Obs. ID} & \colhead{Date} & \colhead{MJD} & \multicolumn2c{Exposure} & \colhead{$S$} & \colhead{Obs. ID} & \colhead{$B$} & \colhead{{Net} Rate} & \colhead{$F_X \times 10^{-13}$} & \colhead{$B$} & \colhead{{Net} Rate} & \colhead{$F_X \times 10^{-13}$} \\ 
\colhead{} & \colhead{(yyyy-mm-dd)} & \colhead{} & \multicolumn2c{(ks)} & \colhead{} & \colhead{} & \colhead{} & \colhead{(10$^{-2}$ count s$^{-1}$)} & \colhead{(
erg cm$^{-2}$ s$^{-1}$)} & \colhead{} & \colhead{(10$^{-2}$ count s$^{-1}$)} & \colhead{(erg cm$^{-2}$ s$^{-1}$)} } 

\decimals
\startdata
172 & 1999-09-05 & 51426.8 & 9.4 & $279\pm{17}$ & 172 & $162\pm{13}$ & 2.73$\pm$0.19 & $3.40_{-0.22}^{+0.22}$ & $321\pm{18}$ & 2.82$\pm$0.19 & $3.51_{-0.22}^{+0.22}$ \\
1857 & 2000-10-04 & 51821.9 & 48.4 & $1531\pm{39}$ & 1857 & $712\pm{27}$ & 2.95$\pm$0.08 & $3.66_{-0.10}^{+0.10}$ & $1420\pm{38}$ & 3.03$\pm$0.08 & $3.74_{-0.10}^{+0.10}$ \\
1038 & 2001-09-19 & 52171.2 & 50.0 & $1491\pm{39}$ & 1038 & $722\pm{27}$ & 2.77$\pm$0.08 & $3.47_{-0.10}^{+0.10}$ & $1460\pm{38}$ & 2.85$\pm$0.08 & $3.55_{-0.10}^{+0.10}$ \\
10227 & 2009-03-20 & 54910.9 & 130.9 & $3538\pm{59}$ & 10227 & $1903\pm{44}$ & 2.49$\pm$0.05 & $3.40_{-0.06}^{+0.06}$ & $4220\pm{65}$ & 2.56$\pm$0.05 & $3.47_{-0.06}^{+0.06}$ \\
10229 & 2009-03-24 & 54914.5 & 48.4 & $1315\pm{36}$ & 10229 & $699\pm{26}$ & 2.51$\pm$0.08 & $3.41_{-0.10}^{+0.10}$ & $1528\pm{39}$ & 2.57$\pm$0.08 & $3.49_{-0.10}^{+0.10}$ \\
10892 & 2009-03-26 & 54916.2 & 124.0 & $3440\pm{59}$ & 10892 & $1756\pm{42}$ & 2.57$\pm$0.05 & $3.50_{-0.06}^{+0.06}$ & $4100\pm{64}$ & 2.62$\pm$0.05 & $3.56_{-0.06}^{+0.06}$ \\
10228 & 2009-03-28 & 54918.9 & 129.5 & $3475\pm{59}$ & 10228 & $1831\pm{43}$ & 2.48$\pm$0.05 & $3.37_{-0.06}^{+0.06}$ & $4226\pm{65}$ & 2.53$\pm$0.05 & $3.44_{-0.06}^{+0.06}$ \\
10698 & 2009-03-31 & 54921.7 & 51.4 & $1476\pm{38}$ & 10698 & $766\pm{28}$ & 2.66$\pm$0.08 & $3.62_{-0.10}^{+0.10}$ & $1592\pm{40}$ & 2.73$\pm$0.08 & $3.70_{-0.10}^{+0.10}$ \\
26629 & 2023-11-15 & 60263.1 & 13.2 & $309\pm{18}$ & 26629 & $176\pm{13}$ & 2.14$\pm$0.14 & $3.28_{-0.20}^{+0.20}$ & $355\pm{19}$ & 2.21$\pm$0.14 & $3.37_{-0.20}^{+0.20}$ \\
29062 & 2023-11-18 & 60266.3 & 9.8 & $214\pm{15}$ & 29062 & $112\pm{11}$ & 2.02$\pm$0.16 & $3.09_{-0.23}^{+0.23}$ & $256\pm{16}$ & 2.07$\pm$0.16 & $3.15_{-0.23}^{+0.23}$ \\
29063 & 2023-11-18 & 60266.9 & 9.8 & $225\pm{15}$ & 29063 & $103\pm{10}$ & 2.14$\pm$0.16 & $3.28_{-0.23}^{+0.23}$ & $272\pm{16}$ & 2.17$\pm$0.16 & $3.30_{-0.23}^{+0.23}$ \\
27098 & 2024-08-06 & 60528.0 & 13.6 & $290\pm{17}$ & 27098 & $158\pm{13}$ & 1.96$\pm$0.13 & $3.06_{-0.19}^{+0.20}$ & $331\pm{18}$ & 2.02$\pm$0.13 & $3.13_{-0.19}^{+0.19}$ \\
29499 & 2024-08-06 & 60528.6 & 13.8 & $314\pm{18}$ & 29499 & $198\pm{14}$ & 2.07$\pm$0.14 & $3.22_{-0.20}^{+0.20}$ & $364\pm{19}$ & 2.15$\pm$0.14 & $3.34_{-0.20}^{+0.20}$ \\
\enddata

\tablecomments{{$S$ and $B$ denote the total counts from the source and background regions, respectively.} Case I and II have the same source extraction region but different background extraction regions, {while the corresponding ratios of background to source region sizes are 6.9 and 21.7 for Case I and II, respectively.} See Section~\ref{sec:observation} for details. {All uncertainties quoted correspond to 1$\sigma$ unless indicated otherwise.}}
\end{splitdeluxetable*}

\section{Data analysis and results} \label{sec:results}

With the determined $\teff$ values of the Cas A NS, we were able to 
measure 
the $\teff$ decline rate of the Cas A NS.
We fitted the $\teff$ values as a function of time using a linear regression law {(in a log scale): 
\begin{equation}
    \log T(t) = \log T_0 - s \log t/t_0,
\end{equation}
where $t$ is the NS age {in years}, $t_0=330$~yr corresponds to MJD=55500 {(near the midpoint of the time interval)}, $T_0 \equiv T(t_0)$, and $s$ is the cooling slope.
We then calculated the Cas A NS cooling rate over 10 years ($C_{10}$) using:
\begin{equation}
    C_{10}\,[\%] = \left(1 - \frac{T(t={340})}{T(t={330})} \right) \times 100.
\end{equation}}

\begin{figure*}
    \centering
    \includegraphics[width=0.495\linewidth]{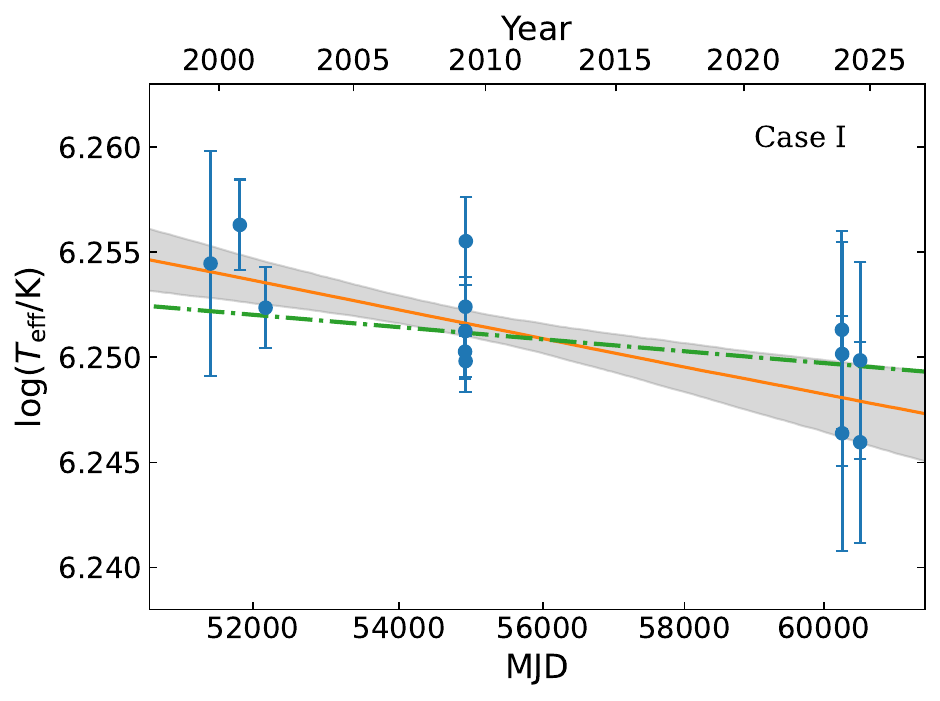}
    \includegraphics[width=0.495\linewidth]{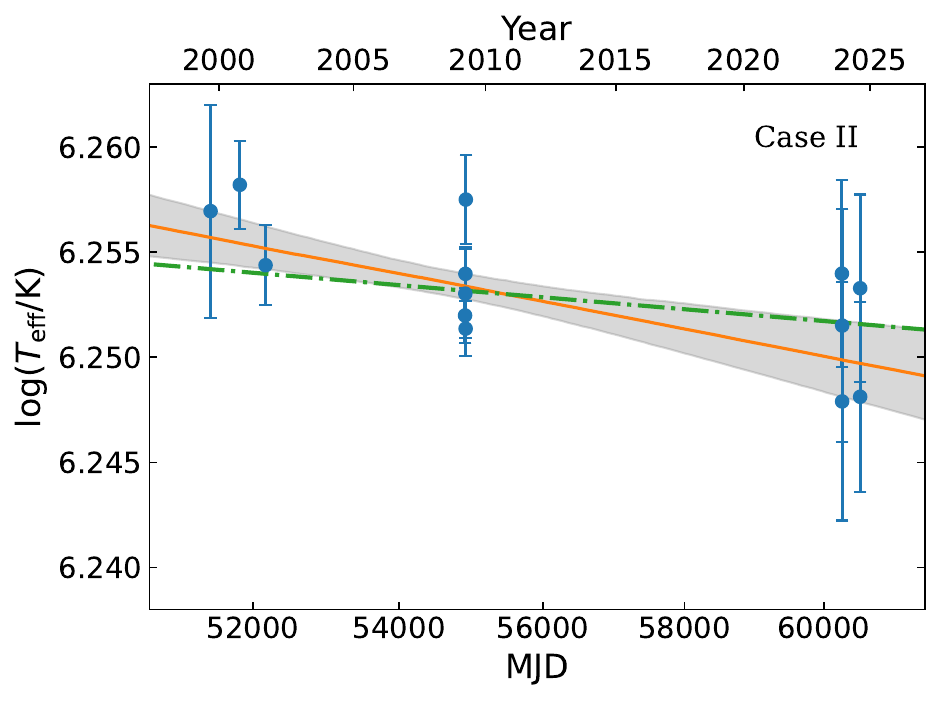}
    \caption{Effective temperatures ($\teff$) of the Cas A NS as a function of time for Case I (left) and II (right) {with \nh fixed}. In each panel, the orange solid line shows the best fit using a linear regression law, whereas the gray shaded region shows the 1$\sigma$ interval of the best fit (see Section~\ref{sec:results} for more details). {For comparison, the green {dot-dashed} line shows the {predicted} standard NS cooling with the cooling slope $s=0.08$.}}
    \label{fig:fitting}
\end{figure*}

{We again adopted the UltraNest package for the line fitting.}
We employed uniform priors for $s$ and $\log T_0$ in the ranges [$-$100,100] and [6.0, 6.5], respectively, for both Case I and II.
The best-fit values of $s$ and $\log T_0$ are $0.19^{+0.09}_{-0.09}$ and $6.251^{+0.001}_{-0.001}$ for Case I {($\chi^2_\nu \approx 0.84$ at posterior median values, $\nu = 11$)}, and $0.18^{+0.08}_{-0.08}$ and $6.253^{+0.001}_{-0.001}$ for Case II {($\chi^2_\nu \approx 1.03$, $\nu = 11$)}, respectively (see Figure~\ref{fig:fitting}).
Consequently, the indicated cooling rates are $0.57_{-0.27}^{+0.26}\%$ and $0.55_{-0.25}^{+0.24}\%$ per decade for Case I and II, respectively.
The cooling rates of the Cas A NS found in this work 
{agree} with the rate reported by \citet{Elshamouty2013} using the same instrument ($1.0\% \pm 0.7\%$ per decade), but with higher significance, which is expected given that the time span in this study is more than twice as long as that in \citet{Elshamouty2013}.

\begin{figure*}
    \centering
    \includegraphics[width=0.495\linewidth]{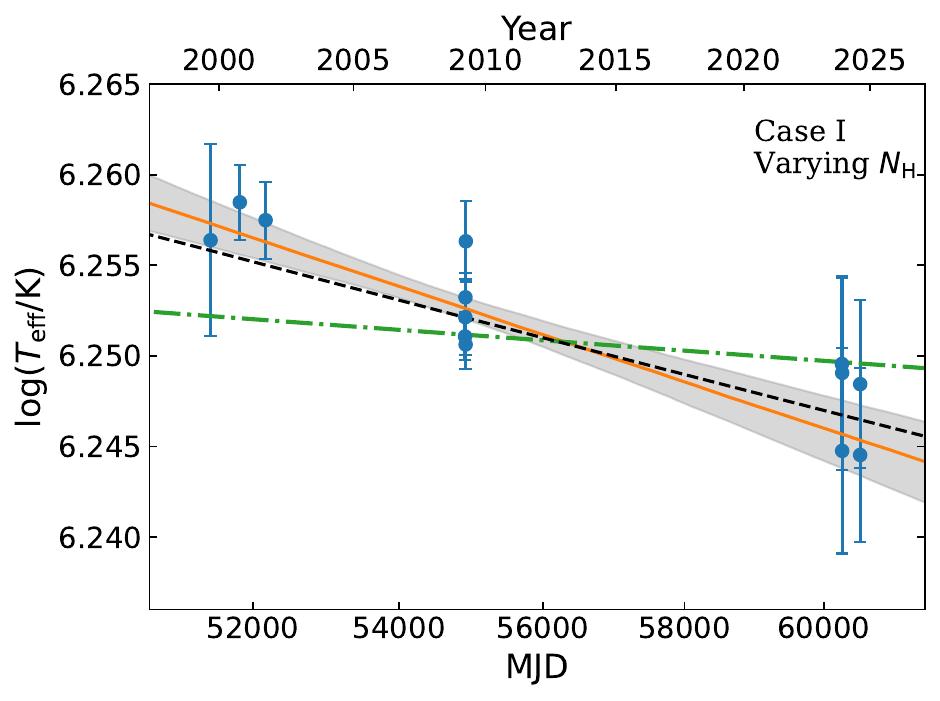}
    \includegraphics[width=0.495\linewidth]{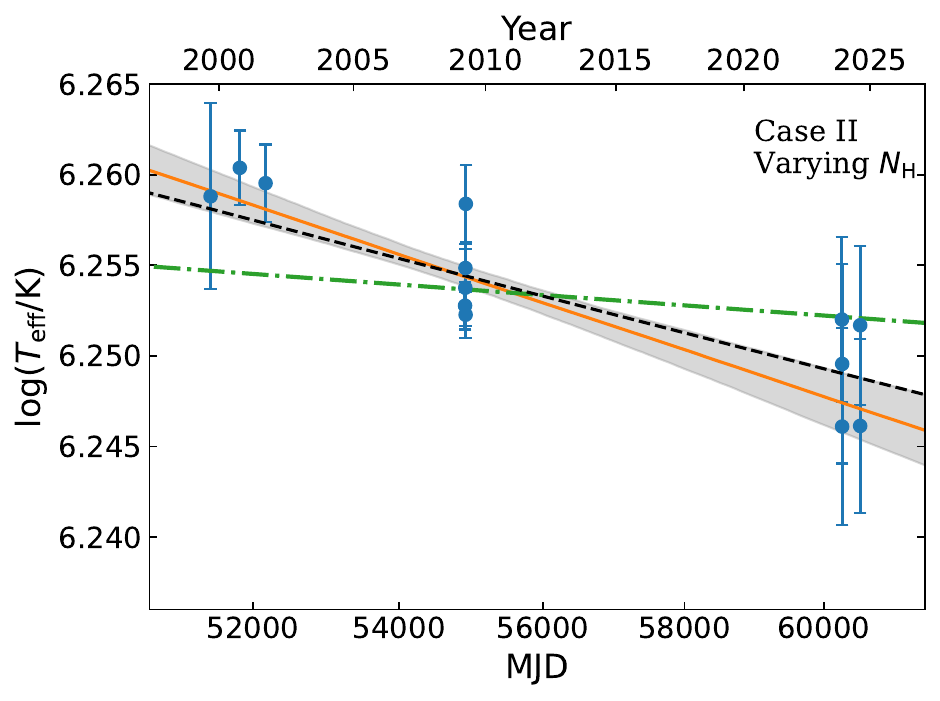}
    \caption{Effective temperatures ($\teff$) of the Cas A NS as a function of time for Case I (left) and II (right), with \nh variations taken into account. In each panel, the orange solid line shows the best fit using a linear regression law, whereas the gray shaded region shows the 1$\sigma$ interval of the best fit. {For comparison, the green {dot-dashed} line shows the {predicted} standard NS cooling with the cooling slope $s=0.08$. The black dashed line represents the {example of simulated} cooling due to the neutron Cooper pair breaking and formation process (see Section~\ref{sec:discussion} for more details).}}
    \label{fig:fitting_varied_nh}
\end{figure*}

Moreover, it is possible that the \nh towards the Cas A NS varies over time (see, e.g., \citealt{Ho2021,Shternin2023}). 
To account for potential \nh variations, we adopted the best-fit \nh values from \citet[Table 2]{Shternin2023}, {that are measured based on ACIS datasets.}
By comparing the observation dates in \citet{Shternin2023} with those in this work, we selected the \nh values closest in time to our observations. {The HRC observations were  within 9 months of an ACIS observation, except the 2024 data which was 4 years after the latest ACIS observations. This makes the $N_H$ assumption for the last HRC datapoints rather uncertain.}
{We then performed the fitting 
again adopting these \nh values (see Table~\ref{tab:Teff}).}
{For Case I, the best-fit values of $s$ and $\log T_0$ are $0.37^{+0.09}_{-0.08}$ and $6.252^{+0.001}_{-0.001}$, respectively {($\chi^2_\nu \approx 0.83$, $\nu = 11$)}, corresponding to a cooling rate of $1.11_{-0.28}^{+0.25}$\% per decade, while for Case II, the best-fit $s$ and $\log T_0$ are $0.36^{+0.09}_{-0.09}$ and $6.254^{+0.001}_{-0.001}$, respectively {($\chi^2_\nu \approx 1.04$, $\nu = 11$)}, leading to a cooling rate of $1.09_{-0.27}^{+0.26}$\% per decade (Figure~\ref{fig:fitting_varied_nh}), {which are larger than the rates derived assuming a fixed \nh}.}

{Alternatively, the change in the measured {net} energy fluxes can also provide evidence of the temperature decline of Cas A NS.
By fitting a line to the measured energy fluxes (see Table~\ref{tab:obs})\footnote{{Note that the net energy fluxes listed in Table~\ref{tab:obs} 
can be seen as estimates of the absorbed fluxes from Cas~A NS with varying \nh}.} for both cases,
we found the flux decline rates over 10 years are
$4.7\% \pm 1.5\%$ {($\chi^2_\nu \approx 0.81$, $\nu = 11$)} and $4.6 \% \pm1.4 \%$ {($\chi^2_\nu \approx 0.81$, $\nu = 11$)}
for Case I and II, respectively.
Again, these fitted decline rates are consistent with the ones in \citet{Elshamouty2013}, but with higher significance.}

\begin{deluxetable*}{lcc|ccc}

\tabletypesize{\small}


\tablecaption{Best-fit effective temperatures ($\teff$) with fixed and varying \nh values for Case I and Case II.\label{tab:Teff}}

\tablenum{2}

\tablehead{\colhead{} & \multicolumn2c{$N_{\rm H}=1.656\times10^{22}~{\rm cm}^{-2}$} & \multicolumn3c{Varing \nh} \\
\cline{2-3} \cline{4-6}
\colhead{Obs. ID} & \colhead{$\log T_{\rm eff, Case~I}$} & \colhead{$\log T_{\rm eff, Case~II}$} & \colhead{\nh} & \colhead{$\log T_{\rm eff, Case~I}$} & \colhead{$\log T_{\rm eff, Case~II}$} \\ 
\colhead{} & \colhead{} & \colhead{} & \colhead{($10^{22}~{\rm cm}^{-2}$)} & \colhead{} & \colhead{} } 

\startdata
172 & $6.254^{+0.005}_{-0.006}$ & $6.257^{+0.005}_{-0.005}$ & 1.69 & $6.256^{+0.006}_{-0.005}$ & $6.259^{+0.005}_{-0.005}$ \\
1857 & $6.256^{+0.002}_{-0.002}$ & $6.258^{+0.002}_{-0.002}$ & 1.69 & $6.258^{+0.002}_{-0.002}$ & $6.260^{+0.002}_{-0.002}$ \\
1038 & $6.252^{+0.002}_{-0.002}$ & $6.254^{+0.002}_{-0.002}$ & 1.74 & $6.257^{+0.002}_{-0.002}$ & $6.260^{+0.002}_{-0.002}$ \\
10227 & $6.250^{+0.001}_{-0.001}$ & $6.252^{+0.001}_{-0.001}$ & 1.67 & $6.251^{+0.001}_{-0.001}$ & $6.253^{+0.001}_{-0.001}$ \\
10229 & $6.251^{+0.002}_{-0.002}$ & $6.253^{+0.002}_{-0.002}$ & 1.67 & $6.252^{+0.002}_{-0.002}$ & $6.254^{+0.002}_{-0.002}$ \\
10892 & $6.252^{+0.001}_{-0.001}$ & $6.254^{+0.001}_{-0.001}$ & 1.67 & $6.253^{+0.001}_{-0.001}$ & $6.255^{+0.001}_{-0.001}$ \\
10228 & $6.250^{+0.001}_{-0.001}$ & $6.251^{+0.001}_{-0.001}$ & 1.67 & $6.251^{+0.001}_{-0.001}$ & $6.252^{+0.001}_{-0.001}$ \\
10698 & $6.256^{+0.002}_{-0.002}$ & $6.258^{+0.002}_{-0.002}$ & 1.67 & $6.256^{+0.002}_{-0.002}$ & $6.258^{+0.002}_{-0.002}$ \\
26629 & $6.251^{+0.004}_{-0.005}$ & $6.254^{+0.005}_{-0.004}$ & 1.63 & $6.250^{+0.005}_{-0.005}$ & $6.252^{+0.005}_{-0.004}$ \\
29062 & $6.246^{+0.006}_{-0.006}$ & $6.248^{+0.006}_{-0.006}$ & 1.63 & $6.245^{+0.006}_{-0.006}$ & $6.246^{+0.005}_{-0.006}$ \\
29063 & $6.250^{+0.005}_{-0.005}$ & $6.252^{+0.005}_{-0.006}$ & 1.63 & $6.249^{+0.005}_{-0.006}$ & $6.250^{+0.005}_{-0.006}$ \\
27098 & $6.246^{+0.004}_{-0.005}$ & $6.248^{+0.004}_{-0.005}$ & 1.63 & $6.245^{+0.005}_{-0.005}$ & $6.246^{+0.005}_{-0.005}$ \\
29499 & $6.250^{+0.004}_{-0.005}$ & $6.253^{+0.005}_{-0.004}$ & 1.63 & $6.248^{+0.004}_{-0.005}$ & $6.252^{+0.005}_{-0.004}$ \\
\enddata


\tablecomments{The \nh values are obtained from \citet[Table 2]{Shternin2023}, which were fitted using different ACIS observations.}


\end{deluxetable*}


\section{Discussion}
\label{sec:discussion}
The most conservative HRC result obtained for the fixed \nh reveals moderately significant cooling at $2\sigma$ level with the cooling exponent of $s=0.19\pm 0.09$. This value does not contradict the standard slow cooling that predicts $s\approx 0.08$ without PBF emission \citep[e.g,][]{Yakovlev2004,Ho2009}. That is, a one-sided probability of $s>0.08$ is only 89\%. When the information on \nh variations is included, we obtain $s=0.36\pm 0.09$, so the cooling significance increases to 4$\sigma$ above no cooling and 3$\sigma$ above the standard slow cooling. The reason for this increase is clear; the decreasing trend in absorption (see Table~\ref{tab:Teff}) requires a stronger decrease of the unabsorbed source flux. {It is important to note} that the analysis of ACIS-S data showed that the fit quality with variable \nh is considerably higher than that for fits with \nh fixed. 
This conclusion is based on the ACIS-S spectral information and may be altered 
{by}
uncalibrated wavelength-dependent systematics {(e.g.\ due to variations of the contamination layer) beyond the best current calibration model \citep{Plucinsky2022}}. 

For both variable and fixed \nh cases, the Cas A NS cooling as measured with HRC-S disagrees at a 3$\sigma$ level with those measured with ACIS-S ($s=0.66^{+0.09}_{-0.07}$ and $0.53^{+0.07}_{-0.05}$ for variable and fixed \nh, respectively, see \citealt{Shternin2023}) and is significantly weaker. This indicates that some systematic mis-calibration is still present between the two instruments. The investigation of the cause of this inconsistency is beyond the scope of the present report and requires further attention. We note, however, that this discrepancy cannot be attributed to the pileup issues, since 
ACIS-S observations
taken in the subarray mode alone, which does not suffer from the pileup, currently 
suggest
even stronger cooling \citep{Posselt2022,Shternin2023}.

Let us now assume that the real Cas A NS cooling rate corresponds to the rate measured with HRC-S under the variable \nh assumption and that the neutron PBF process is responsible for it. This slower cooling does not significantly modify previous estimates for the maximal critical temperature of the neutron superfluid, $T_{Cn\mathrm{max}}$, since, according to the detailed analysis \citep{Shternin2021,Shternin2023}, the latter estimates scale as $s^{1/5}$. Therefore, an alteration of $s$ by a factor of two leads to a modification of $T_{Cn\mathrm{max}}$ only of the order of $15\%$ which is much smaller than uncertainties related to the specific model of the star (its mass, radius, and equation of state) and position of the critical temperature profile maximum within the core. The slower cooling, however, eliminates the tension between the observed and theoretically possible cooling revealed in the previous works based on the ACIS-S data \citep[e.g.,][]{Shternin2011,Shternin2023}. Indeed, the calculations of the neutrino emission accompanying the Cooper pair formation performed with proper account for the response of the condensate in vector and axial-vector channels \citep{Leinson2010} have shown that the emissivity is reduced 
by a factor of $q=0.19$ compared with the results of calculations 
obtained neglecting the condensate response \citep[e.g][]{Yakovlev2001}. This suppression has made it challenging to successfully explain 
{the previous cooling constraints reported by e.g., \citet{Ho2021,Shternin2023}} via the PBF process.

Since the spectral analysis was performed for specific values of NS mass and radius, there is no need to employ the full machinery of the {equation of state and superfluid} model-independent method developed in \citet{Shternin2021}. Instead, we perform cooling simulations with a relatively standard microphysics input to illustrate that the cooling due to the PBF process  in a neutron triplet superfluid is consistent with HRC-S data.
In Figure~\ref{fig:fitting_varied_nh} with the black dashed line, we show an example of such simulations.
Specifically, we used the cooling code described in \citet{Gnedin2001}. We used a $M=1.6\,{\rm M}_\odot$ NS model with cold-catalyzed crust and nucleon core with BSk~24 equation of state \citep{BSk24.2018}.
We used the critical profile for the singlet proton superfluidity from \citet{CCDK1993} denoted by CDDK in \citet{Ho2015} and the critical profile for the triplet neutron superfluidity from \citet{TTav2004} denoted by TTav in \citet{Ho2015}.  The maximum of the TTav critical temperature profile in our model corresponds to $\rho=4\times 10^{14}$~g~cm$^{-3}$. The heat blanketing envelope was assumed to contain a negligible amount of light elements (see detailed discussion in \citealt{Shternin2021,Wijngaarden2019}). The PBF emission strength corresponding to the results of \citet{Leinson2010} is used (i.e. $q=0.19$). By rescaling the maximal critical temperature in the core {with respect to the theoretically calculated maximum of TTav profile, while keeping its shape intact}, it is possible to make the cooling curve pass through the observational points. The black dashed line in Figure~\ref{fig:fitting_varied_nh}  corresponds to $T_{Cn\mathrm{max}}=6.7\times 10^8$~K and $\widetilde{T}_{Cn\mathrm{max}}=4.9\times 10^8$~K, where $\widetilde{T}_{Cn\mathrm{max}}$ is the redshifted maximal critical temperature as seen by a distant observer\footnote{It is the redshifted temperature that is approximately constant in the core of the cooling neutron star \citep[e.g.][]{Yakovlev2004}.}.
{This estimate is within the range obtained from the ACIS-S data analysis.} 
This example shows that weaker cooling observed with HRC-S can be easily explained by the reduced PBF emission from neutron triplet superfluid; moreover, there is room for faster cooling in model calculations provided by increasing the sizes of proton and neutron superfluid regions.


\section{Conclusions}
\label{sec:conclusions}

In this work, we analyzed newly obtained \chandra HRC-S observations as well as archival HRC-S datasets to verify the rapid cooling of the Cas A NS. 
For each observation, we determined the surface temperature of the Cas A NS by simulating spectra and matching them to the observed count rate, where a non-magnetized carbon atmosphere NS model with NS mass and radius of 1.60~M$_\sun$ and 12.6 km was assumed.
With a data time span of about 25 years, {we found that,
if the hydrogen column density is assumed to be fixed at \nh$=1.656\times10^{22}$~cm$^{-2}$, the cooling rates of the Cas A NS are $0.57^{+0.26}_{-0.27}\%$ and $0.55^{+0.24}_{-0.25}\%$ per decade, respectively, depending on the background extraction regions.
If the {ACIS-inferred} variation of the \nh values towards Cas A NS is taken into account, the surface temperature decline rates are found to be 
$1.11^{+0.25}_{-0.28}\%$ and $1.09^{+0.26}_{-0.27}\%$, per decade {(though we do not have an accurate ACIS $N_H$ measurement close in time to the 2024 HRC data)}.}
We note that our measured temperature declines are 
smaller than those found by \citet{Shternin2023}, 1.6$\pm$0.2 or 2.2$\pm$0.3 \% per decade {for fixed or varying \nh, respectively}, or by \citet{Posselt2022}, 1.5$\pm$0.2 or 2.3$\pm$0.3 \% per decade, with independent ACIS datasets. 

{Crucially, our results provide an independent 
{measurement}
of the rapid 
{decline of the absorbed flux from the Cas A NS, at $>3\sigma$ confidence.}
{Adopting the PBF model for the rapid cooling,} the maximal critical temperature of the neutron superfluid, $T_{Cn\mathrm{max}}$, inferred from our work is consistent with the predictions based on ACIS-S data \citep[e.g.][]{Shternin2023}. Moreover, the weaker cooling rate of the Cas A NS found using our HRC-S datasets, compared to that derived from ACIS data, alleviates the tension between the observed and theoretically expected cooling, and can be well explained by the reduced PBF emission from neutron triplet superfluidity.}


\begin{acknowledgments}
{We thank Gregory Sivakoff and George Pavlov for useful conversation. }
JZ is supported by China Scholarship Council (CSC), File No. 202108180023. 
CH is supported by NSERC Discovery Grant RGPIN-2023-04264.
WCGH acknowledges support provided by NASA through Chandra Award Number GO3-24046X issued by the Chandra X-ray Observatory Center (CXC), which is operated by the Smithsonian Astrophysical Observatory for and on behalf of the NASA under contract NAS8-03060. DJP also acknowledges support from the Chandra X-ray Center under NASA contract NAS8-03060.
This research has made use of  software provided by the CXC in the application packages {\sc ciao}, {\sc sherpa}, and {\sc ds9}.
This paper employs a list of Chandra datasets, obtained by the Chandra X-ray Observatory, contained in the Chandra Data Collection~\dataset[DOI: 379]{https://doi.org/10.25574/cdc.379}.
This research has made use of NASA's Astrophysics Data System.
\end{acknowledgments}

%

\vspace{5mm}
\facility{CXO(HRC-S)}


\software{{\sc ciao} \citep{Fruscione2006},  
        {\sc ds9} \citep{Joye2003},
          {\sc sherpa} \citep{Freeman2001,Doe2007},
          UltraNest \citep{Buchner2021},
          Matplotlib \citep{Hunter2007},
          NumPy \citep{harris2020array}
          }





\bibliography{ref}{}

\begin{thebibliography}{}
\expandafter\ifx\csname natexlab\endcsname\relax\def\natexlab#1{#1}\fi
\providecommand{\url}[1]{\href{#1}{#1}}
\providecommand{\dodoi}[1]{doi:~\href{http://doi.org/#1}{\nolinkurl{#1}}}
\providecommand{\doeprint}[1]{\href{http://ascl.net/#1}{\nolinkurl{http://ascl.net/#1}}}
\providecommand{\doarXiv}[1]{\href{https://arxiv.org/abs/#1}{\nolinkurl{https://arxiv.org/abs/#1}}}

\bibitem[{{Alarie} {et~al.}(2014){Alarie}, {Bilodeau}, \& {Drissen}}]{Alarie2014}
{Alarie}, A., {Bilodeau}, A., \& {Drissen}, L. 2014, \mnras, 441, 2996, \dodoi{10.1093/mnras/stu774}

\bibitem[{{Alford} \& {Halpern}(2023)}]{Alford2023}
{Alford}, J.~A.~J., \& {Halpern}, J.~P. 2023, \apj, 944, 36, \dodoi{10.3847/1538-4357/acaf55}

\bibitem[{{Blaschke} {et~al.}(2012){Blaschke}, {Grigorian}, {Voskresensky}, \& {Weber}}]{Blaschke2012}
{Blaschke}, D., {Grigorian}, H., {Voskresensky}, D.~N., \& {Weber}, F. 2012, \prc, 85, 022802, \dodoi{10.1103/PhysRevC.85.022802}

\bibitem[{{Bonanno} {et~al.}(2014){Bonanno}, {Baldo}, {Burgio}, \& {Urpin}}]{Bonanno2014}
{Bonanno}, A., {Baldo}, M., {Burgio}, G.~F., \& {Urpin}, V. 2014, \aap, 561, L5, \dodoi{10.1051/0004-6361/201322514}

\bibitem[{{Buchner}(2016)}]{Buchner2016}
{Buchner}, J. 2016, Statistics and Computing, 26, 383, \dodoi{10.1007/s11222-014-9512-y}

\bibitem[{{Buchner}(2019)}]{Buchner2019}
---. 2019, \pasp, 131, 108005, \dodoi{10.1088/1538-3873/aae7fc}

\bibitem[{{Buchner}(2021)}]{Buchner2021}
---. 2021, The Journal of Open Source Software, 6, 3001, \dodoi{10.21105/joss.03001}

\bibitem[{{Chang} {et~al.}(2010){Chang}, {Bildsten}, \& {Arras}}]{Chang2010}
{Chang}, P., {Bildsten}, L., \& {Arras}, P. 2010, \apj, 723, 719, \dodoi{10.1088/0004-637X/723/1/719}

\bibitem[{{Chen} {et~al.}(1993){Chen}, {Clark}, {Dav{\'e}}, \& {Khodel}}]{CCDK1993}
{Chen}, J.~M.~C., {Clark}, J.~W., {Dav{\'e}}, R.~D., \& {Khodel}, V.~V. 1993, \nphysa, 555, 59, \dodoi{10.1016/0375-9474(93)90314-N}

\bibitem[{{Doe} {et~al.}(2007){Doe}, {Nguyen}, {Stawarz}, {Refsdal}, {Siemiginowska}, \& et~al.}]{Doe2007}
{Doe}, S., {Nguyen}, D., {Stawarz}, C., {et~al.} 2007, in Astronomical Society of the Pacific Conference Series, Vol. 376, Astronomical Data Analysis Software and Systems XVI, ed. R.~A. {Shaw}, F.~{Hill}, \& D.~J. {Bell}, 543

\bibitem[{{Elshamouty} {et~al.}(2013){Elshamouty}, {Heinke}, {Sivakoff}, {Ho}, {Shternin}, {Yakovlev}, {Patnaude}, \& {David}}]{Elshamouty2013}
{Elshamouty}, K.~G., {Heinke}, C.~O., {Sivakoff}, G.~R., {et~al.} 2013, \apj, 777, 22, \dodoi{10.1088/0004-637X/777/1/22}

\bibitem[{{Fesen} {et~al.}(2006){Fesen}, {Hammell}, {Morse}, {Chevalier}, {Borkowski}, {Dopita}, {Gerardy}, {Lawrence}, {Raymond}, \& {van den Bergh}}]{Fesen2006}
{Fesen}, R.~A., {Hammell}, M.~C., {Morse}, J., {et~al.} 2006, ApJ, 645, 283, \dodoi{10.1086/504254}

\bibitem[{{Flowers} {et~al.}(1976){Flowers}, {Ruderman}, \& {Sutherland}}]{Flowers1976}
{Flowers}, E., {Ruderman}, M., \& {Sutherland}, P. 1976, ApJ, 205, 541, \dodoi{10.1086/154308}

\bibitem[{{Freeman} {et~al.}(2001){Freeman}, {Doe}, \& {Siemiginowska}}]{Freeman2001}
{Freeman}, P., {Doe}, S., \& {Siemiginowska}, A. 2001, in Society of Photo-Optical Instrumentation Engineers (SPIE) Conference Series, Vol. 4477, Astronomical Data Analysis, ed. J.-L. {Starck} \& F.~D. {Murtagh}, 76--87, \dodoi{10.1117/12.447161}

\bibitem[{{Fruscione} {et~al.}(2006){Fruscione}, {McDowell}, {Allen}, {Brickhouse}, {Burke}, \& et~al.}]{Fruscione2006}
{Fruscione}, A., {McDowell}, J.~C., {Allen}, G.~E., {et~al.} 2006, in Society of Photo-Optical Instrumentation Engineers (SPIE) Conference Series, Vol. 6270, Observatory Operations: Strategies, Processes, and Systems, ed. D.~R. {Silva} \& R.~E. {Doxsey}, 62701V, \dodoi{10.1117/12.671760}

\bibitem[{{Gnedin} {et~al.}(2001){Gnedin}, {Yakovlev}, \& {Potekhin}}]{Gnedin2001}
{Gnedin}, O.~Y., {Yakovlev}, D.~G., \& {Potekhin}, A.~Y. 2001, \mnras, 324, 725, \dodoi{10.1046/j.1365-8711.2001.04359.x}

\bibitem[{{Halpern} \& {Gotthelf}(2010)}]{Halpern2010}
{Halpern}, J.~P., \& {Gotthelf}, E.~V. 2010, \apj, 709, 436, \dodoi{10.1088/0004-637X/709/1/436}

\bibitem[{Harris {et~al.}(2020)Harris, Millman, van~der Walt, Gommers, Virtanen, Cournapeau, Wieser, Taylor, Berg, Smith, Kern, Picus, Hoyer, van Kerkwijk, Brett, Haldane, del R{\'{i}}o, Wiebe, Peterson, G{\'{e}}rard-Marchant, Sheppard, Reddy, Weckesser, Abbasi, Gohlke, \& Oliphant}]{harris2020array}
Harris, C.~R., Millman, K.~J., van~der Walt, S.~J., {et~al.} 2020, Nature, 585, 357, \dodoi{10.1038/s41586-020-2649-2}

\bibitem[{{Heinke} \& {Ho}(2010)}]{Heinke2010}
{Heinke}, C.~O., \& {Ho}, W. C.~G. 2010, ApJL, 719, L167, \dodoi{10.1088/2041-8205/719/2/L167}

\bibitem[{{Ho} {et~al.}(2015){Ho}, {Elshamouty}, {Heinke}, \& {Potekhin}}]{Ho2015}
{Ho}, W. C.~G., {Elshamouty}, K.~G., {Heinke}, C.~O., \& {Potekhin}, A.~Y. 2015, \prc, 91, 015806, \dodoi{10.1103/PhysRevC.91.015806}

\bibitem[{{Ho} \& {Heinke}(2009)}]{Ho2009}
{Ho}, W. C.~G., \& {Heinke}, C.~O. 2009, \nat, 462, 71, \dodoi{10.1038/nature08525}

\bibitem[{{Ho} {et~al.}(2021){Ho}, {Zhao}, {Heinke}, {Kaplan}, {Shternin}, \& {Wijngaarden}}]{Ho2021}
{Ho}, W. C.~G., {Zhao}, Y., {Heinke}, C.~O., {et~al.} 2021, \mnras, 506, 5015, \dodoi{10.1093/mnras/stab2081}

\bibitem[{Hunter(2007)}]{Hunter2007}
Hunter, J.~D. 2007, Computing in Science \& Engineering, 9, 90, \dodoi{10.1109/MCSE.2007.55}

\bibitem[{{Joye} \& {Mandel}(2003)}]{Joye2003}
{Joye}, W.~A., \& {Mandel}, E. 2003, in Astronomical Society of the Pacific Conference Series, Vol. 295, Astronomical Data Analysis Software and Systems XII, ed. H.~E. {Payne}, R.~I. {Jedrzejewski}, \& R.~N. {Hook}, 489

\bibitem[{{Koehn} {et~al.}(2025){Koehn}, {Rose}, {Pang}, {Somasundaram}, {Reed}, {Tews}, {Abac}, {Komoltsev}, {Kunert}, {Kurkela}, {Coughlin}, {Healy}, \& {Dietrich}}]{Koehn2025}
{Koehn}, H., {Rose}, H., {Pang}, P. T.~H., {et~al.} 2025, PhRvX, 021014, \dodoi{10.1103/PhysRevX.15.021014}

\bibitem[{{Leinson}(2010)}]{Leinson2010}
{Leinson}, L.~B. 2010, \prc, 81, 025501, \dodoi{10.1103/PhysRevC.81.025501}

\bibitem[{{Leinson}(2022)}]{Leinson2022}
---. 2022, \mnras, 511, 5843, \dodoi{10.1093/mnras/stac448}

\bibitem[{{McLaughlin} {et~al.}(2001){McLaughlin}, {Cordes}, {Deshpande}, {Gaensler}, {Hankins}, {Kaspi}, \& {Kern}}]{McLaughlin2001}
{McLaughlin}, M.~A., {Cordes}, J.~M., {Deshpande}, A.~A., {et~al.} 2001, \apjl, 547, L41, \dodoi{10.1086/318891}

\bibitem[{{Murray} {et~al.}(2002){Murray}, {Ransom}, {Juda}, {Hwang}, \& {Holt}}]{Murray2002}
{Murray}, S.~S., {Ransom}, S.~M., {Juda}, M., {Hwang}, U., \& {Holt}, S.~S. 2002, \apj, 566, 1039, \dodoi{10.1086/338224}

\bibitem[{{Page} {et~al.}(2011){Page}, {Prakash}, {Lattimer}, \& {Steiner}}]{Page2011}
{Page}, D., {Prakash}, M., {Lattimer}, J.~M., \& {Steiner}, A.~W. 2011, PhRvL, 106, 081101, \dodoi{10.1103/PhysRevLett.106.081101}

\bibitem[{{Pearson} {et~al.}(2018){Pearson}, {Chamel}, {Potekhin}, {Fantina}, {Ducoin}, {Dutta}, \& {Goriely}}]{BSk24.2018}
{Pearson}, J.~M., {Chamel}, N., {Potekhin}, A.~Y., {et~al.} 2018, \mnras, 481, 2994, \dodoi{10.1093/mnras/sty2413}

\bibitem[{{Plucinsky} {et~al.}(2022){Plucinsky}, {Bogdan}, \& {Marshall}}]{Plucinsky2022}
{Plucinsky}, P.~P., {Bogdan}, A., \& {Marshall}, H.~L. 2022, in Society of Photo-Optical Instrumentation Engineers (SPIE) Conference Series, Vol. 12181, Space Telescopes and Instrumentation 2022: Ultraviolet to Gamma Ray, ed. J.-W.~A. {den Herder}, S.~{Nikzad}, \& K.~{Nakazawa}, 121816X, \dodoi{10.1117/12.2630193}

\bibitem[{{Posselt} \& {Pavlov}(2018)}]{Posselt2018}
{Posselt}, B., \& {Pavlov}, G.~G. 2018, \apj, 864, 135, \dodoi{10.3847/1538-4357/aad7fc}

\bibitem[{{Posselt} \& {Pavlov}(2022)}]{Posselt2022}
---. 2022, \apj, 932, 83, \dodoi{10.3847/1538-4357/ac6dca}

\bibitem[{{Posselt} {et~al.}(2013){Posselt}, {Pavlov}, {Suleimanov}, \& {Kargaltsev}}]{Posselt2013}
{Posselt}, B., {Pavlov}, G.~G., {Suleimanov}, V., \& {Kargaltsev}, O. 2013, \apj, 779, 186, \dodoi{10.1088/0004-637X/779/2/186}

\bibitem[{Potekhin \& Yakovlev(2025)}]{Potekhin2025}
Potekhin, A., \& Yakovlev, D. 2025, Journal of High Energy Astrophysics, 49, 100441, \dodoi{10.1016/j.jheap.2025.100441}

\bibitem[{{Predehl} {et~al.}(2003){Predehl}, {Costantini}, {Hasinger}, \& {Tanaka}}]{Predehl2003}
{Predehl}, P., {Costantini}, E., {Hasinger}, G., \& {Tanaka}, Y. 2003, Astronomische Nachrichten, 324, 73, \dodoi{10.1002/asna.200310019}

\bibitem[{{Ransom}(2002)}]{Ransom2002}
{Ransom}, S.~M. 2002, in Astronomical Society of the Pacific Conference Series, Vol. 271, Neutron Stars in Supernova Remnants, ed. P.~O. {Slane} \& B.~M. {Gaensler}, 361, \dodoi{10.48550/arXiv.astro-ph/0112006}

\bibitem[{{Reed} {et~al.}(1995){Reed}, {Hester}, {Fabian}, \& {Winkler}}]{Reed1995}
{Reed}, J.~E., {Hester}, J.~J., {Fabian}, A.~C., \& {Winkler}, P.~F. 1995, \apj, 440, 706, \dodoi{10.1086/175308}

\bibitem[{{Shternin} {et~al.}(2023){Shternin}, {Ofengeim}, {Heinke}, \& {Ho}}]{Shternin2023}
{Shternin}, P.~S., {Ofengeim}, D.~D., {Heinke}, C.~O., \& {Ho}, W. C.~G. 2023, \mnras, 518, 2775, \dodoi{10.1093/mnras/stac3226}

\bibitem[{{Shternin} {et~al.}(2021){Shternin}, {Ofengeim}, {Ho}, {Heinke}, {Wijngaarden}, \& {Patnaude}}]{Shternin2021}
{Shternin}, P.~S., {Ofengeim}, D.~D., {Ho}, W. C.~G., {et~al.} 2021, \mnras, 506, 709, \dodoi{10.1093/mnras/stab1695}

\bibitem[{{Shternin} {et~al.}(2011){Shternin}, {Yakovlev}, {Heinke}, {Ho}, \& {Patnaude}}]{Shternin2011}
{Shternin}, P.~S., {Yakovlev}, D.~G., {Heinke}, C.~O., {Ho}, W. C.~G., \& {Patnaude}, D.~J. 2011, MNRAS, 412, L108, \dodoi{10.1111/j.1745-3933.2011.01015.x}

\bibitem[{{Takatsuka} \& {Tamagaki}(2004)}]{TTav2004}
{Takatsuka}, T., \& {Tamagaki}, R. 2004, Progress of Theoretical Physics, 112, 37, \dodoi{10.1143/PTP.112.37}

\bibitem[{{Tananbaum}(1999)}]{Tananbaum1999}
{Tananbaum}, H. 1999, \iaucirc, 7246, 1

\bibitem[{{Verner} {et~al.}(1996){Verner}, {Ferland}, {Korista}, \& {Yakovlev}}]{Verner1996}
{Verner}, D.~A., {Ferland}, G.~J., {Korista}, K.~T., \& {Yakovlev}, D.~G. 1996, \apj, 465, 487, \dodoi{10.1086/177435}

\bibitem[{{Wijngaarden} {et~al.}(2019){Wijngaarden}, {Ho}, {Chang}, {Heinke}, {Page}, {Beznogov}, \& {Patnaude}}]{Wijngaarden2019}
{Wijngaarden}, M.~J.~P., {Ho}, W. C.~G., {Chang}, P., {et~al.} 2019, \mnras, 484, 974, \dodoi{10.1093/mnras/stz042}

\bibitem[{{Wilms} {et~al.}(2000){Wilms}, {Allen}, \& {McCray}}]{Wilms2000}
{Wilms}, J., {Allen}, A., \& {McCray}, R. 2000, \apj, 542, 914, \dodoi{10.1086/317016}

\bibitem[{{Wu} {et~al.}(2021){Wu}, {Pires}, {Schwope}, {Xiao}, {Yan}, \& {Ji}}]{Wu2021}
{Wu}, Q., {Pires}, A.~M., {Schwope}, A., {et~al.} 2021, Research in Astronomy and Astrophysics, 21, 294, \dodoi{10.1088/1674-4527/21/11/294}

\bibitem[{{Yakovlev} {et~al.}(2011){Yakovlev}, {Ho}, {Shternin}, {Heinke}, \& {Potekhin}}]{Yakovlev2011}
{Yakovlev}, D.~G., {Ho}, W. C.~G., {Shternin}, P.~S., {Heinke}, C.~O., \& {Potekhin}, A.~Y. 2011, \mnras, 411, 1977, \dodoi{10.1111/j.1365-2966.2010.17827.x}

\bibitem[{{Yakovlev} {et~al.}(2001){Yakovlev}, {Kaminker}, {Gnedin}, \& {Haensel}}]{Yakovlev2001}
{Yakovlev}, D.~G., {Kaminker}, A.~D., {Gnedin}, O.~Y., \& {Haensel}, P. 2001, \physrep, 354, 1, \dodoi{10.1016/S0370-1573(00)00131-9}

\bibitem[{{Yakovlev} \& {Pethick}(2004)}]{Yakovlev2004}
{Yakovlev}, D.~G., \& {Pethick}, C.~J. 2004, ARA\&A, 42, 169, \dodoi{10.1146/annurev.astro.42.053102.134013}

\bibitem[{{Yang} {et~al.}(2011){Yang}, {Pi}, \& {Zheng}}]{Yang2011}
{Yang}, S.-H., {Pi}, C.-M., \& {Zheng}, X.-P. 2011, \apjl, 735, L29, \dodoi{10.1088/2041-8205/735/2/L29}

\end{thebibliography}
\bibliographystyle{aasjournal}


\end{CJK*}
\end{document}